\documentclass[aps,twocolumn,showpacs,prl,preprintnumbers,amsmath,amssymb,superscriptaddress]{revtex4-2}
\usepackage{epsf}
\usepackage{graphicx}
\usepackage{sidecap}
 \usepackage{soul}
 \usepackage{array}
 \usepackage{amsmath}
 \usepackage{amssymb}
 \usepackage{dcolumn}
 \usepackage{epstopdf}
 \usepackage{bm}
 \usepackage{amsmath}
 
\usepackage{gensymb}
\usepackage{color}
\usepackage{hyperref}
\usepackage{soul}
\sethlcolor{green}
\usepackage{bbding}
\usepackage{sidecap}
\usepackage{float}

\hypersetup{
    colorlinks=true,
    citecolor=red,
    linkcolor=red,
    filecolor=blue,   
    urlcolor=blue,
}
 
\def \beq {\begin{equation}}
\def \eeq {\end{equation}}
\def\bibsection{\refname}
\renewcommand{\refname}{\noindent\textbf{References}}
\begin{document}
\title{Low-lying Electronic Structure of Rare-Earth Based Topological Nodal Line Semimetal Candidate DySbTe}
\author{Nathan~Valadez}\affiliation {Department of Physics, University of Central Florida, Orlando, Florida 32816, USA}

\author{Iftakhar~Bin~Elius}\affiliation {Department of Physics, University of Central Florida, Orlando, Florida 32816, USA}
\author{Dante~James}\affiliation {Department of Physics, University of Central Florida, Orlando, Florida 32816, USA}
\author{Peter~Radanovich}\affiliation {Department of Physics, University of Central Florida, Orlando, Florida 32816, USA}
\author{Tetiana~Romanova}\affiliation {Institute of Low Temperature and Structure Research, Polish Academy of Sciences, 50-950 Wrocław, Poland}
\author{Sami~Elgalal}\affiliation {Institute of Low Temperature and Structure Research, Polish Academy of Sciences, 50-950 Wrocław, Poland}
\author{Grzegorz~Chajewski}\affiliation {Institute of Low Temperature and Structure Research, Polish Academy of Sciences, 50-950 Wrocław, Poland}
\author{Florie Mesple}\affiliation {Department of Physics, University of Washington, Seattle, Washington, 98195, USA}
\author{Ellis Thompson}\affiliation {Department of Physics, University of Washington, Seattle, Washington, 98195, USA}
\author{Keng Tou Chu}\affiliation {Department of Physics, University of Washington, Seattle, Washington, 98195, USA}
\author{Matthew Yankowitz}\affiliation {Department of Physics, University of Washington, Seattle, Washington, 98195, USA}
\affiliation {Department of Materials Science and Engineering, University of Washington, Seattle, Washington, 98195, USA}
\author{Andrzej~Ptok}\affiliation {Institute of Nuclear Physics, Polish Academy of Sciences, W. E. Radzikowskiego 152, PL-31342 Krak\'ow, Poland}
\author{Dariusz~Kaczorowski}\affiliation {Institute of Low Temperature and Structure Research, Polish Academy of Sciences, 50-950 Wrocław, Poland}
\author{Madhab Neupane} \thanks{Corresponding author:\href{mailto:madhab.neupane@ucf.edu}{madhab.neupane@ucf.edu}}\affiliation{Department of Physics, University of Central Florida, Orlando, Florida 32816, USA}

\date{\today}

\begin{abstract}
Lanthanide ($Ln$) based $Ln$SbTe materials have garnered significant attention due to rich interplay of long-range magnetic ordering and topological properties, driven by unique crystalline symmetry, 4$f$ electron interactions, and pronounced spin-orbit coupling (SOC) effects. DySbTe, as a heavier lanthanide-based member of the $Ln$SbTe family, stands out with its SOC and larger on-site interactions on its 4$f$ electrons, which arise due to the heavier Dy element. Here, we present a comprehensive study on the low-temperature bulk physical properties and the electronic structure of DySbTe using magnetic susceptibility, heat capacity, and electrical resistivity measurements, along with high-resolution angle-resolved photoemission spectroscopy (ARPES), scanning tunneling microscopy and spectroscopy (STM/S), and density functional theory calculations. Our thermodynamic measurements revealed an antiferromagnetic ordering below $T_{\rm N}$ = 7.45 K and a subsequent magnetic phase transition at $T_{\rm N1}$ = 7.15 K. Our transport studies indicate a semimetallic behavior with unusual feature in the ordered state. Our ARPES measurements revealed a diamond-shaped Fermi pocket centered at the $\overline{\Gamma}$ point, with band features that evolve distinctly across various binding energies. STM/S results indicate a minimum in the density of states at around 100 meV below the Fermi level, and ARPES measurements reveal a significant gap present around the $\overline{\text{X}}$ point, differentiating DySbTe from other $Ln$SbTe compounds. These findings enhance our understanding of the SOC effects on the electronic structure and topological properties in the $Ln$SbTe family, highlighting DySbTe as a promising candidate for exploring the interplay between topology and magnetism. 
\end{abstract}

\maketitle  
\indent The rapidly expanding field of topological quantum materials (TQMs) has undergone significant advancements since the initial discovery of three-dimensional (3D) topological insulators~\cite{col_hasan, Qi_TI, TopologicalInsulators_Ortmann}, semimetals~\cite{wang, Neupane2014, Yang2014, Xu2015, Weylding, Soluyanov2015, Physweyl, Burkov_WeylSemimetalTopoInsulMulti, Fang_NLSM_2015, node_surface_liang, Neupane_16, bian2016topological, hosen_ZrSiX, nodal_loop, Armitagereview, weyl_nodal_loop}, and superconductors~\cite{ShouCheng, Sato_2017}. 
The identification of various topological semimetals, including Dirac semimetals~\cite{wang, Neupane2014, Yang2014}, Weyl semimetals~\cite{Xu2015, Weylding, Soluyanov2015, Physweyl, Burkov_WeylSemimetalTopoInsulMulti}, and nodal-line semimetals (NLSMs)~\cite{Fang_NLSM_2015, node_surface_liang, Neupane_16, bian2016topological, hosen_ZrSiX, nodal_loop, Armitagereview, weyl_nodal_loop}, has been driven by a surge of both theoretical and experimental research. The identification of nodal-line topological states in ZrSiS has significantly expanded interest in ZrSiS-type materials~\cite{Neupane_16, schoop2016dirac, Lv_ZrSiSSingleCrystal}. These materials, characterized by nonsymmorphic topological fermions from the square-net plane of group-IV elements~\cite{Neupane_16, schoop2016dirac}, exhibit remarkable electronic properties such as, unconventional magnetotransport behavior~\cite{Singha_MagnetoresistanceZrSiS}, flat optical conductivity~\cite{ZrSiS_OpticalConductivity}, and enhanced effective mass~\cite{ZrSiS_MassEnhancement} which has further motivated the exploration of additional members in this material family. These studies of nodal line structures in this family also underscore the need for further exploration into the effects of electron hybridization, including interactions influenced by magnetism, 4$f$-electron correlations, and strong spin-orbit coupling (SOC). Factors such as these uniquely affect the nodal line features and demand deeper investigation to understand their impact on the topological and magnetic characteristics of these materials. \\
\begin{figure*} 
\includegraphics[width=18cm]{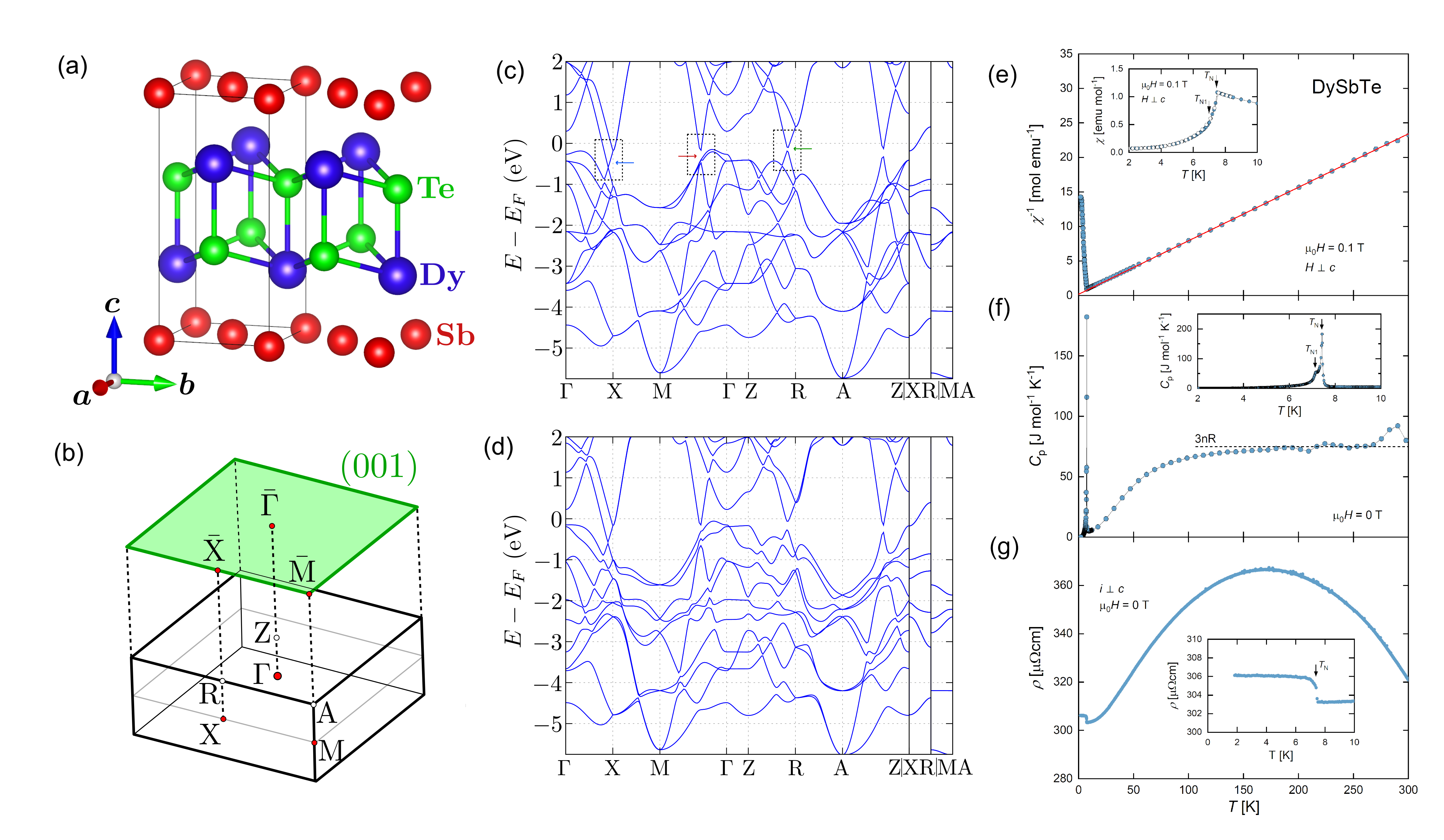} 
   \vspace{-1ex}
	\caption{Crystalline, electronic structures, and bulk properties of single-crystalline DySbTe. (a) Crystallographic unit cell. (b) Bulk Brillouin zone with the projected surface Brillouin zone marked with high symmetry points. Calculated bulk band structures along high symmetry directions (c) without and with (d) SOC taken into consideration. Temperature dependencies of (e) the inverse magnetic susceptibility, (f) the specific heat, and (g) the electrical resistivity measured with the magnetic field and electric current directions specified in the panels. The insets of panels (e,f) present the low-temperature data of (e) the magnetic susceptibility measured in zero-field-cooled (filled symbols) and field-cooled (open symbols) regimes,  (f) the specific heat, and (g) the electrical resistivity. The red straight line in panel (e) marks the Curie-Weiss behavior described in the text. The horizontal dashed line in panel (f) represents the Dulong-Petit limit}
\label{fig1}
\end{figure*}
\indent The $Ln$SbTe systems, part of the ZrSiS-type family of materials, have attracted significant interest due to the less-studied Sb-square net, potential magnetic ordering from $Ln$ magnetic moments, and correlation effects from the 4$f$ electrons of $Ln$ elements. Nonsymmorphic symmetry, involving fractional lattice translations with operations like screw axes or glide planes, has been associated with symmetry-protected topology~\cite{nonsymmomorphic_zhao, nonsymmorphic_michel, nonsymmorphic_kane}. However, materials with non-accidental nodal lines without SOC are limited~\cite{Fang_NLSM_2015, Neupane_16, nonsymmomorphic_Yang, nonsymmomorphic_zhao, schoop2016dirac}. Recent studies proposed continuous Dirac nodal points, validated by observing nodal line phases in ZrSiS and other MZX (M$=$ Transition metals, Z$=$ Si, Ge, Sb, Sn, X$=$S, Se, Te) materials with PbFCl-type structures~\cite{Neupane_16, schoop2016dirac}. Replacing transition metals with rare earth elements alters topological properties due to 4$f$ electron correlations and magnetism. Generally, lighter lanthanide compounds exhibit gapless nodal lines, as seen in LaSbTe, where the nodal lines along R\textminus X and A\textminus M directions remain robust against SOC due to nonsymmorphic symmetry~\cite{La}. In contrast, CeSbTe exhibits meta-magnetic transitions, charge density waves (CDW), and weak Kondo effects~\cite{CeSbTe_cao, CeSbTe_Lv, CeSbTe_peng}. SmSbTe displayed multiple Dirac nodes alongside Kondo effects and electronic correlations~\cite{CeSbSe_Chen, SmSbTe_sabin}, while PrSbTe and NdSbTe revealed gapless nodal lines and diamond-shaped structures~\cite{PrSbTe}, with the latter showing magnetic studies indicating metamagnetic transitions, Kondo localization~\cite{NdSbTeEandMProp_Pandey}, and enhanced correlations~\cite{Nd111}. GdSbTe exhibits ZrSiS-like nodal line characteristics at different energy positions~\cite{GdSbTe_hosen}, with early work revealing nodal-line and AFM Dirac states protected by broken time-reversal and rotoinversion symmetries~\cite{GdSbTe_hosen}. While extensive studies have focused on lighter lanthanide-based compounds, the stronger SOC and more localized 4$f$-electron interactions in heavier elements, like Dy, have yet to reveal significant changes in their electronic structures, leaving much of their potential properties underexplored. The SOC gap in HoSbTe at the $\overline{\text{X}}$ point suggests it as a weak topological insulator, with magnetism-driven band structure changes near $4$K~\cite{HoSbTe_arpes}. HoSbTe also shows SOC-induced gaps along specific high-symmetry directions, driving a Lifshitz transition that leads to a Fermi surface (FS) with four electron pockets centered at the $\overline{\text{X}}$ point~\cite{HoSbTe_arpes, HoSbTe_shumiya, HoSBTe_yang}. In contrast, lighter $Ln$SbTe retains a gapless nodal line even with SOC~\cite{LaSbTe_transport, Schoop}, highlighting Dy as a transition between gapless and gapped topological states across the series. \\
\indent DySbTe is another representative of the $Ln$SbTe family that was characterised by means of thermodynamic and transport measurements as an AFM semimetal~\cite{Gao_DySbTe}. Recent neutron diffraction experiment revealed on DySbTe competing magnetic phases within the AFM state~\cite{plokhikh_Ln}. In this paper, we report on our magnetic susceptibility, heat capacity, and electrical resistivity measurements performed on high-quality single crystals of DySbTe. The ARPES measurements, conducted in the paramagnetic (PM) phase at $15$~K, reveal persistent nodal line features along several high-symmetry directions, and measurements are further complemented by first-principles band structure calculations. Notably, DySbTe marks a significant case within the $Ln$SbTe family as it transitions from gapless to gapped states, showing gapped band crossings along the $\overline{\text{X}}-\overline{\Gamma}-\overline{\text{X}}$ and $\overline{\text{M}}-\overline{\Gamma}-\overline{\text{M}}$ directions. Additionally, the nodal line along the bulk X\textminus R direction is also gapped. These findings are consistent with first-principles calculations and provide key insights into the electronic structure and topological states of DySbTe, establishing a strong foundation for understanding the interplay between topology and $Ln$ elements in the $Ln$SbTe family of quantum materials. \\
\begin{figure} 
	\includegraphics[width=7cm]{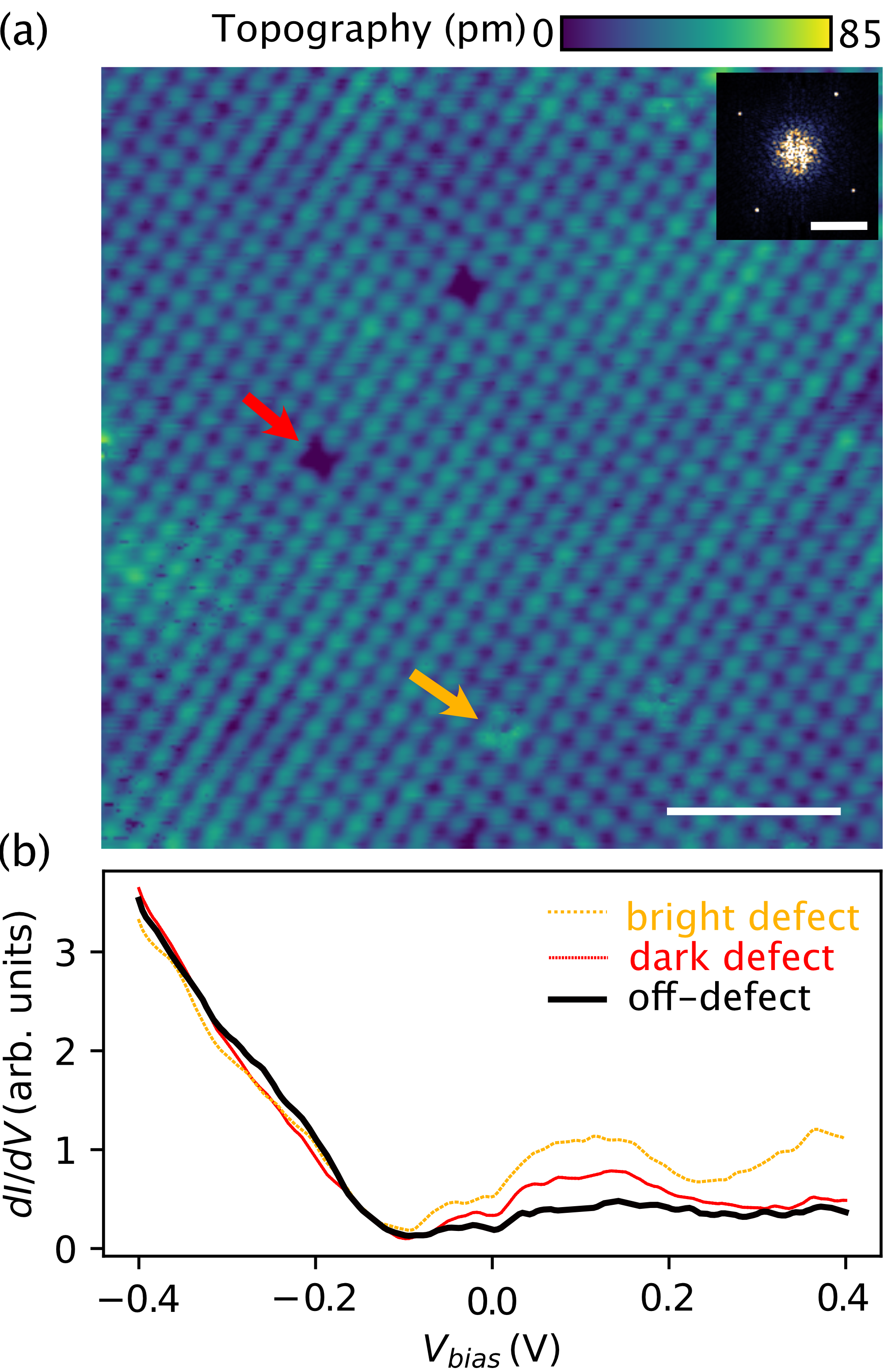} 
    \vspace{-1ex}
	\caption{STM topography and STS spectra of DySbTe. (a) STM topography taken at setpoint conditions $(V_{bias}, I_{set})=(0.3~\text{V}, 150~\text{pA})$. In addition to the atomic lattice, dark and bright atomic defects are marked by the red and yellow arrows, respectively. The scale bar is 2~nm.  The inset shows an FFT of the topograph. Bright is high and dark is low intensity, and the scale bar is 2~nm$^{-1}$. (b) STS spectra taken on the pristine lattice, as well as on the bright and dark defects.}
\label{fig_new}
\end{figure}
\begin{figure*}
	\includegraphics[width=18cm]{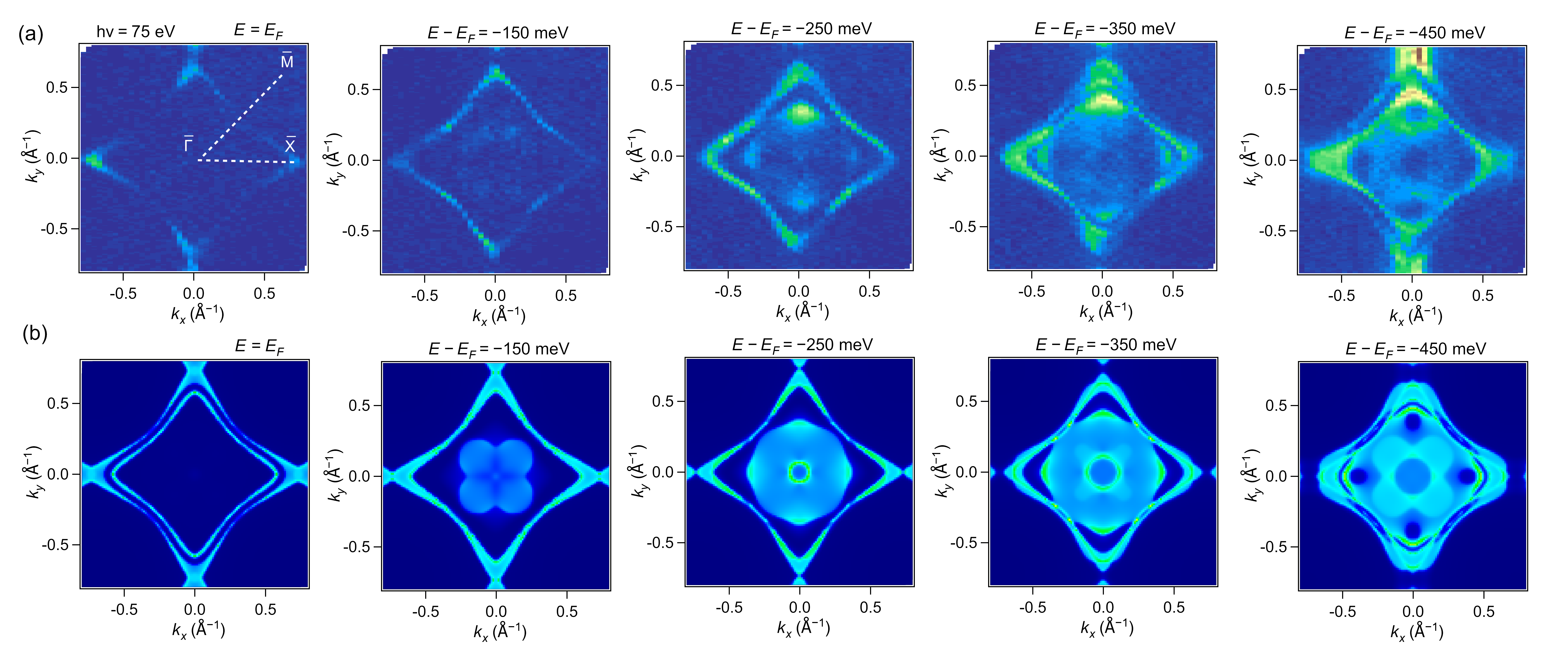} 
    \vspace{-1ex}
	\caption{Plots of the Fermi map and constant energy contours in DySbTe. (a) ARPES measured Fermi surface (first leftmost panel ) and constant energy contours with various binding energies as noted on the top of each plot. 
    (b) Theoretically obtained respective FS and energy contours. 
    Experimental data was collected at SSRL endstation 5-2 at a temperature of $15$~K.}
\label{fig3}
\end{figure*}
\indent Single crystals of DySbTe were grown from Sb flux as described in the Supplementary Material (SM) to this manuscript~\cite{SupplementalMat}. Their quality and chemical composition were proved by standard techniques (see SM~\cite{SupplementalMat}). Measurements of the magnetic susceptibility, heat capacity and electrical resistivity were carried out through the usage of Quantum Design PPMS-9 and MPMS-XL. STM and ARPES experiments were conducted and analysed as described in SM~\cite{SupplementalMat}. First-principles calculations were performed within the framework of density functional theory (DFT), using the Vienna Ab initio Simulation Package (VASP), based on projector augmented wave (PAW) potentials (See SM~\cite{SupplementalMat} for additional information)~\cite{kresse.hafner.94,kresse.furthmuller.96,kresse.joubert.99,blochl.94}. \\
\begin{figure*}
	\includegraphics[width=18cm]{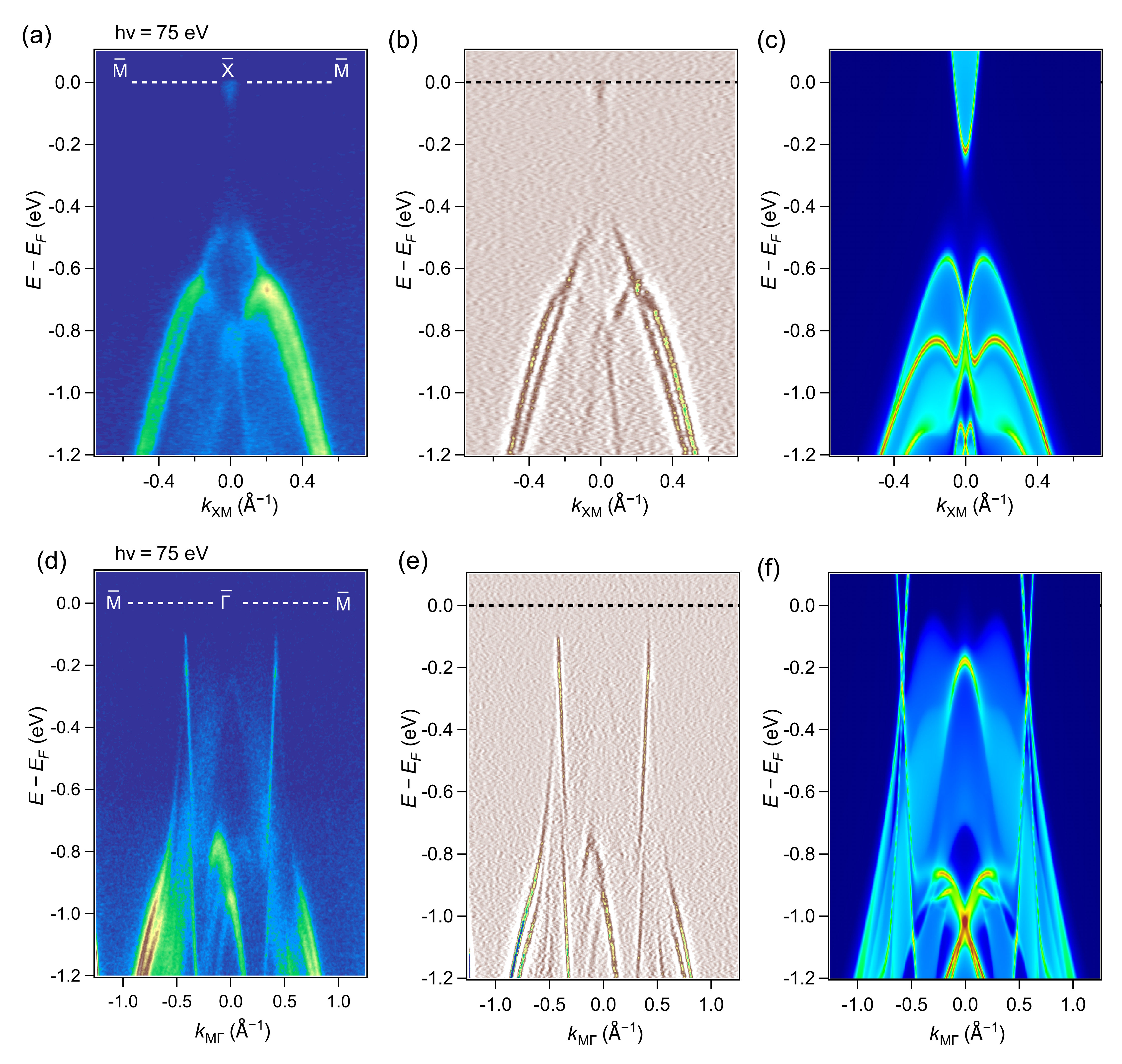} 
    \vspace{-1ex}
	\caption{Band structure of DySbTe along the $\overline{\text{M}}-\overline{\text{X}}-\overline{\text{M}}$ and $\overline{\text{M}}-\overline{\Gamma}-\overline{\text{M}}$. (a) ARPES measured band dispersion along the $\overline{\text{M}}-\overline{\text{X}}-\overline{\text{M}}$ direction with a measured photon energy of 75 eV. (b) Second derivative plot of (a). (c) Calculated surface electronic structure along $\overline{\text{M}}-\overline{\text{X}}-\overline{\text{M}}$ direction. (d) ARPES measured band dispersion along the $\overline{\text{M}}-\overline{\Gamma}-\overline{\text{M}}$ direction with a measured photon energy of 75 eV. (e) Second derivative plot of (d). (f) Calculated surface electronic structure along $\overline{\text{M}}-\overline{\Gamma}-\overline{\text{M}}$ direction. Experimental data was collected at SSRL endstation 5-2 at a temperature of 15K.}
\label{fig3}
\end{figure*}
\begin{figure*}
	\includegraphics[width=18cm]{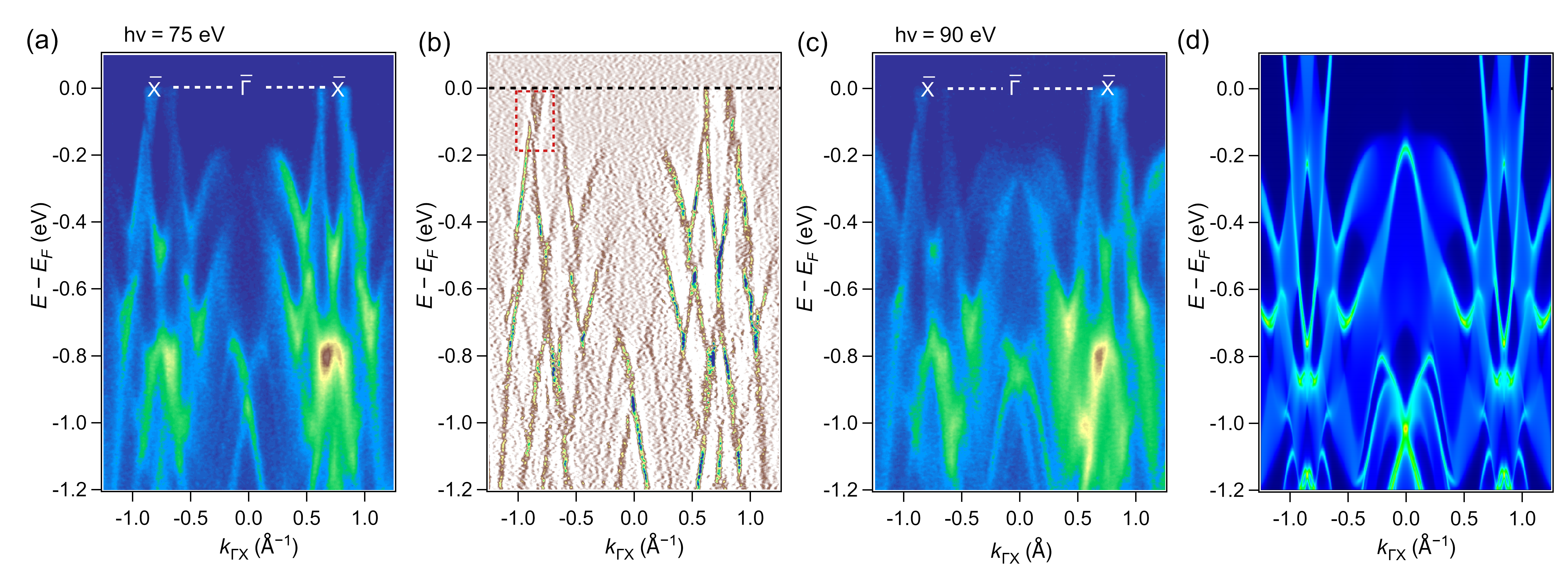} 
    \vspace{-1ex}
	\caption{Band structure of DySbTe along the $\overline{\text{X}}$-$\overline{\Gamma}$-$\overline{\text{X}}$. (a) ARPES measured band dispersion along the $\overline{\text{X}}-\overline{\Gamma}-\overline{\text{X}}$ direction with a measured photon energy of 75 eV. (b) Second derivative plot of (a). (c) ARPES measured band dispersion along the $\overline{\text{X}}-\overline{\Gamma}-\overline{\text{X}}$ with a different photon energy noted on top of the plot. (d) Calculated surface electronic structure along $\overline{\text{X}}-\overline{\Gamma}-\overline{\text{X}}$ direction. Experimental data was collected at SSRL endstation 5-2 at a temperature of $15$~K.}
\label{fig4}
\end{figure*}
\indent The tetragonal crystal structure of DySbTe is shown in Fig.1(a). The compound is isostructural with other $Ln$SbTe materials which crystallize within a $P4/nmm$ (No.~129) nonsymmorphic space group. The lattice parameters reported in the literature are $a = b = 4.240$~\AA\ and $c = 9.168$~\AA~\cite{Gao_DySbTe}, and our theoretical crystal structure optimization yielded similar values: $a = b = 4.286$~\AA\ and $c = 9.235$~\AA. In the crystallographic unit cell, the Sb atom occupies the Wyckoff position $2b$ (1/4,3/4,1/2), while the Dy and Te atoms are located at the $2c$ (3/4,3/4,$z$) sites, where $z = 0.2743$ and $0.6238$ for Dy and Te, respectively. The Dy-Te double layers are situated between the Sb atoms arranged in a square lattice configuration. These layers weakly bonded to neighboring sheets via van der Waals interactions, resulting in a natural (001) cleaving plane.\\
\indent Fig.1(b) illustrates the three-dimensional bulk Brillouin zone (BZ) and its projection along the (001) surface, where ARPES experiments were conducted, showing the corresponding high symmetry points. The calculated bulk band structures, without and with the inclusion of SOC, are shown in Figs.1(c) and1(d). In Fig.1(c), locations of high interest are indicated by a black box and a possible gap formation due to the inclusion of SOC is indicated by a blue arrow, and other gapped features are indicated by a red and green arrow. \\ 
\indent The bulk thermodynamic and electrical transport properties of the DySbTe single crystals are presented in Fig.1(e-g). Magnetic measurements performed in a magnetic field of 0.1 T applied in the tetragonal plane of the compound corroborated the AFM order at low temperatures, first reported in Refs.~\cite{Gao_DySbTe,plokhikh_Ln}. However, the Neel temperature $T_{\rm N}$ = 7.45 K, defined at the maximum in the temperature dependence of the molar magnetic susceptibility (see the inset to Fig.1(e)) is slightly higher than the values given in the literature (7 K~\cite{Gao_DySbTe}, 6.7 K~\cite{plokhikh_Ln}). Noteworthy is the rapidity of the decrease in the magnetic susceptibility just below $T_{\rm N}$. In the ordered state, the magnetic susceptibility was found nearly independent of the magnetic history of the specimen, in line with the AFM character of the electronic ground state of DySbTe. However, a small change in the slope of $\chi$($T$) measured in the field-cooled regime can be detected near 7.1 K. This latter feature was not observed in the previous studies. \\ 
\indent In the paramagnetic state, the magnetic susceptibility follows the Curie-Weiss law (note the red line in Fig.1(e)) with the effective magnetic moment $\mu_{\rm eff}$ = 10.2 $\mu_{\rm B}$ and the paramagnetic temperature $\theta_{\rm p} = -2.4$ K. The experimental value of $\mu_{\rm eff}$ is fairly close to the theoretical prediction for a free trivalent Dy ion (10.65 $\mu_{\rm B}$), calculated assuming the Russell-Saunders coupling scenario. Some tiny deviation of $\chi^{-1}$($T$) from a straight line, occurring below about 40 K, probably results from crystalline electric field interactions. The negative sign of $\theta_{\rm p}$ complies with the AFM coupling of localized paramagnetic moments. \\
\indent The magnetic phase transition at $T_{\rm N}$ manifests itself as a distinct anomaly in the temperature-dependent specific heat of the compound. As can be inferred from Fig.1(f), the peak in $C_{\rm p}$($T$) is extremely sharp, resembling first-order transitions. Remarkably, another small peak appears in the specific heat data at $T_{\rm N1}$ = 7.15 K (see the inset to Fig.1(f)), which coincides with the results of the magnetic measurements. Above 200 K, $C_{\rm p}$($T$) saturates at a value of 75 J/(mole K), which is equal to the Dulong-Petit limit 3$nR$, where $n$ is the number of atoms in the formula unit and $R$ stands for the gas constant. The observed unevenness of $C_{\rm p}$($T$) above 210 K probably results from the contribution of Apiezon N grease~\cite{Apiezon}, which was quite significant in measuring the heat capacity of a very small DySbTe crystal. \\
\indent As shown in Fig.1(g), the studied crystals of DySbTe were found to exhibit semimetallic character of the electrical transport in terms of the magnitude of the electrical resistivity, its little variation with temperature, and the overall shape of $\rho$($T$). As the temperature is lowered below 300 K, the resistivity first increases, passes through a broad maximum centered near 170 K, and then decreases with further temperature reduction. When approaching $T_{\rm N}$, $\rho$($T$) shows a plateau and then jumps suddenly at the onset of the AFM state. The sharpness of this anomaly corresponds very well with the first-order-like character of the peak in $C_{\rm p}$($T$). Interestingly, $\rho$($T$) flattens out below $T_{\rm N1}$ and the resistivity remains constant down to 2 K. This behavior is rather unusual for AFM materials and may result from the complexity of the magnetic structure of DySbTe that features two propagation vectors and was interpreted in terms of two-phase model~\cite{plokhikh_Ln}.\\ 
\indent It is worth noting that the overall behavior of $\rho$($T$) in the PM region is similar to that found earlier~\cite{Gao_DySbTe}, but the resistivity is about four times smaller than in the previous report. The key difference is in the ordered state, namely, in contrast to our finding, a slight decrease in the resistivity was observed below $T_{\rm N}$ in Ref.~\cite{Gao_DySbTe}. This dissimilarity hints at a possible difference in the stoichiometry of the crystals investigated in the two studies, which is a common feature of the $Ln$SbTe compounds (see for example Ref.~\cite{nonstoichiometry} and references therein).\\
\indent Fig.2(a) shows an STM topograph of DySbTe cleaved in ultra-high vacuum along the (001) plane. A rectangular atomic lattice is visible, consistent with our theoretically predicted crystal structure. Additionally, there are two types of atomic defects present, attributed to Te vacancies (dark, red arrow) and adatoms or substitutions (bright, orange arrow) \cite{bian_visualizing_2022}. Atomic Bragg peaks in the FFT (inset) indicate lattice constants of $a=4.1\pm0.2~\text{\AA}$ and $b=4.0\pm0.2~\text{\AA}$. Fig.2(b) shows STS spectra taken on and away from the defects, displaying a lack of a hard gap within a 800~meV window around the Fermi level ($V_{bias}=0$), consistent with our band structure calculations. \\
\indent To further investigate the compound, we turn to a detailed examination of the electronic band structure of DySbTe through ARPES measurements. In this comprehensive study of the band structure of DySbTe, we present the experimentally obtained constant energy contours (CECs) from ARPES at the Fermi level and various binding energies using an incident photon energy of 75 eV, as shown in 3(a). The leftmost panel of 3(a) reveals the FS characterized by a diamond-shaped Fermi pocket centered around the $\Gamma$ point, a distinctive feature of ZrSiS-type materials. The FS shows small pockets around the $\overline{\text{X}}$ points, with a narrow pocket extending towards neighboring $\overline{\text{X}}$ points to form the diamond shape. Notably, the FS is gapped specifically along the $\overline{\text{M}}-\overline{\Gamma}-\overline{\text{M}}$ direction with a pronounced decrease in spectral intensity, while the corners of the diamond pocket remain ungapped, highlighting areas of persistent electronic states. As the binding energy increases, we observe that the band along the $\overline{\text{M}}-\overline{\Gamma}- \overline{\text{M}}$ direction begins to fill around $-$150 meV. The band features within the CECs become more pronounced with increasing binding energies. At approximately $-$250 meV below the Fermi level, a hole pocket containing a double sheet and features near the $\Gamma$ point begin to form. This evolution continues, with the diamond-shaped pocket developing into a triple sheet band around the $\overline{\text{X}}-\overline{\Gamma}- \overline{\text{X}}$ direction at around $-$350 meV and then reverting to a double sheet nature at around $-$450 meV. These experimental CECs can be compared to the calculated CECs presented in 3(b). At the Fermi energy, our experimental data reveal a gap along the $\overline{\text{M}}-\overline{\Gamma}- \overline{\text{M}}$ direction without the double sheet feature observed in theoretical results. Both the experimental and theoretical results show open, diamond-shaped Fermi pockets; however, the experimental pockets close rapidly at higher binding energies within the Brillouin zone, as opposed to the theoretical calculations.\\
\indent To further investigate the electronic structure of DySbTe, we focus on the dispersion maps along various high symmetry directions. Specifically, Fig.4(a) presents the experimental dispersion cut along the $\overline{\text{M}}-\overline{\text{X}}-\overline{\text{M}}$ direction.  The bands observed along this direction reveal a significant gapped feature at the $\overline{\text{X}}$ point with a band gap around 300 meV, which differs to other lighter members in the $Ln$SbTe family that have little to no gapped feature along this direction~\cite{PrSbTe, Nd111, SmSbTe_sabin, GdSbTe_hosen}. This gap can be further visualized in the second derivative plot shown in Fig.4(b). These experimental observations are compared with the calculated band structure along the $\overline{\text{M}}-\overline{\text{X}}-\overline{\text{M}}$ direction, as depicted in Fig.4(c), which shows good agreement with the experimental data. Next, we explore the band features along the $\overline{\text{M}}-\overline{\Gamma}-\overline{\text{M}}$ direction using an incident photon energy of 75 eV, as shown in Fig.4(d). In this dispersion cut, a very sharp, prominently linearly dispersive band with a high band velocity of around 4.74 eV $\text{\AA}$ is observed. This band extends close to the Fermi level, terminating around 0.1 eV below it. This feature again remains unseen in other lighter $Ln$SbTe members which have their linear dispersive bands extend up to the Fermi level~\cite{PrSbTe, Nd111, SmSbTe_sabin, GdSbTe_hosen}. The second derivative plot in Fig.4(e) highlights this linear band further, showing its movement away from the $\overline{\Gamma}$ point. Additionally, starting around 0.8 eV below the Fermi level, multiple bands with varying band velocities are seen crossing near the $\overline{\Gamma}$ point. These findings are corroborated by the calculated band structure shown in Fig.4(f), which predicts that the bands should extend up to the Fermi level; however, experimental results reveal a gap instead. Despite this discrepancy, the calculated and experimental data are otherwise in good agreement. \\
\indent We then investigated the dispersion map along the $\overline{\text{X}}-\overline{\Gamma}-\overline{\text{X}}$ direction, as presented in Fig.5(a), using an incident photon energy of 75 eV. Similar to other ZrSiS-type materials, we anticipated multiple Dirac crossings along the $\overline{\text{X}}-\overline{\Gamma}- \overline{\text{X}}$ direction. To verify if these Dirac crossings persist in DySbTe, we examined the second derivative plot shown in Fig.5(b). From both Figs.5(a) and (b), we observe linear dispersive bands around the $\overline{\text{X}}$ point, along with a gap located approximately 0.1 eV below the Fermi level (surrounded by a dashed red rectangle in Fig.5(b)). This is consistent with our STS measurements in Fig.2(b), which show a minimum in the d$I$/d$V$ spectroscopy at $V_{bias}\approx -100$~meV. This observation is distinct from other $Ln$SbTe members, where a nodal line crossing typically exists near the Fermi level~\cite{PrSbTe, Nd111, SmSbTe_sabin, GdSbTe_hosen}. Further analysis along the $\overline{\text{X}}-\overline{\Gamma}-\overline{\text{X}}$ direction with an incident photon energy of 90 eV, as shown in Fig.5(c), reveals that the induced gap remains near the $\overline{\text{X}}$ point. These experimental findings are compared to the calculated band structure in Fig.5(d), showing reasonable agreement.\\
\indent In conclusion, we have investigated the electronic structure of DySbTe through high-resolution ARPES and STM/S measurements with complementary DFT calculations. Our experimental data reveal a diamond-shaped Fermi pocket centered around the $\overline{\Gamma}$ point, characteristic of the ZrSiS-type materials with the nonsymmorphic $P4/nmm$ space group, where distinct band features evolve at various binding energies. Notably, STM/S results indicate a minimization of the density of states at around 100 meV below the Fermi level, and we observed, through ARPES, a gap feature along the $\overline{\text{M}}-\overline{\text{X}}-\overline{\text{M}}$ and $\overline{\text{M}}-\overline{\Gamma}-\overline{\text{M}}$ direction with a additional gap located at the $\overline{\text{X}}$ point along the $\overline{\text{X}}-\overline{\Gamma}-\overline{\text{X}}$, distinguishing DySbTe from other lighter $Ln$SbTe members. The low-temperature thermodynamic and electrical transport measurements, combined with STS, confirmed an AFM character of the electronic ground state in DySbTe and a semimetallic nature of the compound. All our findings provide a deeper understanding of the influence of SOC on the electronic structure and topological properties within the $Ln$SbTe family. \\
\indent M.N. acknowledges the support from the National Science Foundation (NSF) CAREER Award No. DMR-1847962, and the NSF Partnerships for Research and Education in Materials (PREM) Grant No. DMR-2424976. M.Y. acknowledges support from the National Science Foundation (NSF) MRSEC 2308979. T.R., S.E., G.C., and D.K. are supported by the National Science Centre (Poland) under research grant 2021/41/B/ST3/01141. A.P. acknowledges the support by National Science Centre (NCN, Poland) under Project No. 2021/43/B/ST3/02166. This research utilized resources of the Stanford Synchrotron Radiation Lightsource (SSRL), SLAC National Accelerator Laboratory, which is a DOE Office of Science User Facility under Contract No. DE-AC02-76SF00515. We express our gratitude to Dr. Makoto Hashimoto and Dr. Donghui Lu for providing valuable beamline support at SSRL. We acknowledge Milo Sprague and Anup Pradhan Sakhya for their help during the experiment and fruitful discussion during the manuscript preparation. \\
\def\bibsection{\section*{\refname}}

\vspace{2ex}

%

\begin{thebibliography}{60}
\bibitem{col_hasan} M. Z. Hasan and C. L. Kane, Colloquium: Topological insulators, \href{https://link.aps.org/doi/10.1103/RevModPhys.82.3045}{Rev. Mod. Phys. \textbf{82}, 3045 (2010)}.

\bibitem{Qi_TI} X.-L. Qi and S.-C. Zhang, Topological insulators and superconductors, \href{https://link.aps.org/doi/10.1103/RevModPhys.83.1057}{Rev. Mod. Phys. \textbf{83}, 1057 (2011)}.

\bibitem{TopologicalInsulators_Ortmann} F. Ortmann, S. Roche, and S. O. Valenzuela, \textit{Topological Insulators: Fundamentals and Perspectives} (Wiley-VCH, 2015), \href{https://doi.org/10.1002/9783527681594}{DOI:10.1002/9783527681594}.

\bibitem{wang} K. Pandey, R. Basnet, A. Wegner, G. Acharya, M. R. U. Nabi, J. Liu, J. Wang, Y. K. Takahashi, B. Da, and J. Hu, Electronic and magnetic properties of the topological semimetal candidate NdSbTe, \href{https://link.aps.org/doi/10.1103/PhysRevB.101.235161}{Phys. Rev. B \textbf{101}, 235161 (2020)}.

\bibitem{Neupane2014} M. Neupane, S.-Y. Xu, R. Sankar, N. Alidoust, G. Bian, C. Liu, I. Belopolski, T.-R. Chang, H.-T. Jeng, H. Lin, A. Bansil, F. Chou, and M. Z. Hasan, Observation of a three-dimensional topological Dirac semimetal phase in high-mobility Cd$_{3}$As$_{2}$, \href{https://doi.org/10.1038/ncomms4786}{Nat. Commun. \textbf{5}, 3786 (2014)}.

\bibitem{Yang2014} B.-J. Yang and N. Nagaosa, Classification of stable three-dimensional Dirac semimetals with nontrivial topology, \href{https://doi.org/10.1038/ncomms5898}{Nat. Commun. \textbf{5}, 4898 (2014)}.

\bibitem{Xu2015} S.-Y. Xu, I. Belopolski, N. Alidoust, M. Neupane, G. Bian, C. Zhang, R. Sankar, G. Chang, Z. Yuan, C.-C. Lee, S.-M. Huang, H. Zheng, J. Ma, D. S. Sanchez, B. Wang, A. Bansil, F. Chou, P. P. Shibayev, H. Lin, S. Jia, and M. Z. Hasan, Discovery of a Weyl fermion semimetal and topological Fermi arcs, \href{https://doi.org/10.1126/science.aaa9297}{Science \textbf{349}, 613 (2015)}.

\bibitem{Weylding} B. Q. Lv, H. M. Weng, B. B. Fu, X. P. Wang, H. Miao, J. Ma, P. Richard, X. C. Huang, L. X. Zhao, G. F. Chen, Z. Fang, X. Dai, T. Qian, and H. Ding, Experimental Discovery of Weyl Semimetal TaAs, \href{https://link.aps.org/doi/10.1103/PhysRevX.5.031013}{Phys. Rev. X \textbf{5}, 031013 (2015)}.

\bibitem{Soluyanov2015} A. A. Soluyanov, D. Gresch, Z. Wang, Q. Wu, M. Troyer, X. Dai, and B. A. Bernevig, Type-II Weyl semimetals, \href{http://dx.doi.org/10.1038/nature15768}{Nature \textbf{527}, 495 (2015)}.

\bibitem{Physweyl} H. Weng, C. Fang, Z. Fang, B. A. Bernevig, and X. Dai, Weyl Semimetal Phase in Noncentrosymmetric Transition-Metal Monophosphides, \href{https://link.aps.org/doi/10.1103/PhysRevX.5.011029}{Phys. Rev. X \textbf{5}, 011029 (2015)}.

\bibitem{Burkov_WeylSemimetalTopoInsulMulti} A. A. Burkov and L. Balents, Weyl Semimetal in a Topological Insulator Multilayer, \href{https://link.aps.org/doi/10.1103/PhysRevLett.107.127205}{Phys. Rev. Lett. \textbf{107}, 127205 (2011)}

\bibitem{Fang_NLSM_2015} C. Fang, Y. Chen, H.-Y. Kee, and L. Fu, Topological nodal line semimetals with and without spin-orbital coupling, \href{https://link.aps.org/doi/10.1103/PhysRevB.92.081201}{Phys. Rev. B \textbf{92}, 081201 (2015)}

\bibitem{node_surface_liang} Q.-F. Liang, J. Zhou, R. Yu, Z. Wang, and H. Weng, Node-surface and node-line fermions from nonsymmorphic lattice symmetries, \href{https://link.aps.org/doi/10.1103/PhysRevB.93.085427}{Phys. Rev. B \textbf{93}, 085427 (2016)}

\bibitem{Neupane_16} M. Neupane, I. Belopolski, M. M. Hosen, D. S. Sanchez, R. Sankar, M. Szlawska, S.-Y. Xu, D. Klauss, N. Dhakal, P. Maldonado, P. M. Oppeneer, D. Kaczorowski, F. Chou, M. Z. Hasan, and T. Durakiewicz, Observation of topological nodal fermion semimetal phase in $\mathrm{ZrSiS}$, \href{https://link.aps.org/doi/10.1103/PhysRevB.93.201104}{Phys. Rev. B \textbf{93}, 201104 (2016)}

\bibitem{bian2016topological} G. Bian, T.-R. Chang, R. Sankar, S.-Y. Xu, H. Zheng, T. Neupert, C.-K. Chiu, S.-M. Huang, G. Chang, I. Belopolski, et al., Topological nodal-line fermions in spin-orbit metal {PbTaSe$_{2}$}, \href{https://doi.org/10.1038/ncomms10556}{Nat. Commun. \textbf{7}, 10556 (2016)}

\bibitem{hosen_ZrSiX} M. M. Hosen, D. Dimitri, I. Belopolski, P. Maldonado, R. Sankar, N. Dhakal, G. Dhakal, T. Cole, P. M. Oppeneer, D. Kaczorowski, F. Chou, M. Z. Hasan, T. Durakiewicz, and M. Neupane, Tunability of the topological nodal-line semimetal phase in $\mathrm{ZrSi}X$-type materials ($X=\mathrm{S}, \mathrm{Se}, \mathrm{Te}$), \href{https://link.aps.org/doi/10.1103/PhysRevB.95.161101}{Phys. Rev. B \textbf{95}, 161101 (2017)}

\bibitem{nodal_loop} J. Zhan, J. Li, W. Shi, X.-Q. Chen, and Y. Sun, Coexistence of {Weyl} semimetal and {Weyl} nodal loop semimetal phases in a collinear antiferromagnet, \href{https://link.aps.org/doi/10.1103/PhysRevB.107.224402}{Phys. Rev. B \textbf{107}, 224402 (2023)}

\bibitem{Armitagereview} N. P. Armitage, E. J. Mele, and A. Vishwanath, Weyl and Dirac semimetals in three-dimensional solids, \href{https://link.aps.org/doi/10.1103/RevModPhys.90.015001}{Rev. Mod. Phys. \textbf{90}, 015001 (2018)}

\bibitem{weyl_nodal_loop} J. Zhan, J. Li, W. Shi, X.-Q. Chen, and Y. Sun, Coexistence of {Weyl} semimetal and {Weyl} nodal loop semimetal phases in a collinear antiferromagnet, \href{https://link.aps.org/doi/10.1103/PhysRevB.107.224402}{Phys. Rev. B \textbf{107}, 224402 (2023)}

\bibitem{ShouCheng} X.-L. Qi and S.-C. Zhang, Topological insulators and superconductors, \href{https://link.aps.org/doi/10.1103/RevModPhys.83.1057}{Rev. Mod. Phys. \textbf{83}, 1057--1110 (2011)}

\bibitem{Sato_2017} M. Sato and Y. Ando, Topological superconductors: a review, \href{https://doi.org/10.1088/1361-6633/aa6ac7}{Rep. Prog. Phys. \textbf{80}, 076501 (2017)}

\bibitem{schoop2016dirac} L. M. Schoop, M. N. Ali, C. Stra{\ss}er, A. Topp, A. Varykhalov, D. Marchenko, V. Duppel, S. S. Parkin, B. V. Lotsch, and C. R. Ast, Dirac cone protected by non-symmorphic symmetry and three-dimensional {Dirac} line node in {ZrSiS}, \href{https://doi.org/10.1038/ncomms11696}{Nat. Commun. \textbf{7}, 11696 (2016)}

\bibitem{Lv_ZrSiSSingleCrystal} Y.-Y. Lv, B.-B. Zhang, X. Li, S.-H. Yao, Y. B. Chen, J. Zhou, S.-T. Zhang, M.-H. Lu, Y.-F. Chen, Extremely large and significantly anisotropic magnetoresistance in ZrSiS single crystals, \href{https://doi.org/10.1063/1.4953772}{Appl. Phys. Lett. \textbf{108}, 244101 (2016)}

\bibitem{Singha_MagnetoresistanceZrSiS} R. Singha, A. K. Pariari, B. Satpati, and P. Mandal, Large nonsaturating magnetoresistance and signature of nondegenerate Dirac nodes in ZrSiS, \href{https://www.pnas.org/doi/abs/10.1073/pnas.1618004114}{PNAS \textbf{114}, 2468-2473 (2017)}

\bibitem{ZrSiS_OpticalConductivity} M. B. Schilling, L. M. Schoop, B. V. Lotsch, M. Dressel, and A. V. Pronin, Flat Optical Conductivity in ZrSiS due to Two-Dimensional Dirac Bands, \href{https://link.aps.org/doi/10.1103/PhysRevLett.119.187401}{Phys. Rev. Lett. \textbf{119}, 187401 (2017)}.

\bibitem{ZrSiS_MassEnhancement} S. Pezzini, M. R. van Delft, L. M. Schoop, B. V. Lotsch, A. Carrington, M. I. Katsnelson, N. E. Hussey, and S. Wiedmann, Unconventional mass enhancement around the Dirac nodal loop in ZrSiS, \href{https://doi.org/10.1038/nphys4306}{Nat. Phys. \textbf{14}, 178 (2018)}.

\bibitem{nonsymmomorphic_zhao} Y. X. Zhao and A. P. Schnyder, Nonsymmorphic symmetry-required band crossings in topological semimetals, \href{https://link.aps.org/doi/10.1103/PhysRevB.94.195109}{Phys. Rev. B \textbf{94}, 195109 (2016)}.

\bibitem{nonsymmorphic_michel} L. Michel and J. Zak, Connectivity of energy bands in crystals, \href{https://link.aps.org/doi/10.1103/PhysRevB.59.5998}{Phys. Rev. B \textbf{59}, 5998 (1999)}.

\bibitem{nonsymmorphic_kane} S. M. Young and C. L. Kane, Dirac Semimetals in Two Dimensions, \href{https://link.aps.org/doi/10.1103/PhysRevLett.115.126803}{Phys. Rev. Lett. \textbf{115}, 126803 (2015)}.

\bibitem{nonsymmomorphic_Yang} B.-J. Yang, T. A. Bojesen, T. Morimoto, and A. Furusaki, Topological semimetals protected by off-centered symmetries in nonsymmorphic crystals, \href{https://link.aps.org/doi/10.1103/PhysRevB.95.075135}{Phys. Rev. B \textbf{95}, 075135 (2017)}.

\bibitem{La} Y. Wang, Y. Qian, M. Yang, H. Chen, C. Li, Z. Tan, Y. Cai, W. Zhao, S. Gao, Y. Feng, S. Kumar, E. F. Schwier, L. Zhao, H. Weng, Y. Shi, G. Wang, Y. Song, Y. Huang, K. Shimada, Z. Xu, X. J. Zhou, and G. Liu, Spectroscopic evidence for the realization of a genuine topological nodal-line semimetal in LaSbTe, \href{https://link.aps.org/doi/10.1103/PhysRevB.103.125131}{Phys. Rev. B \textbf{103}, 125131 (2021)}.

\bibitem{CeSbTe_cao} L. Y. Cao, M. Yang, L. Wang, Y. Li, B. X. Gao, L. Wang, J. L. Liu, A. F. Fang, Y. G. Shi, and R. Y. Chen, Optical study of the topological materials $Ln\mathrm{SbTe}$ $(Ln=\mathrm{La}, \mathrm{Ce}, \mathrm{Sm}, \mathrm{Gd})$, \href{https://link.aps.org/doi/10.1103/PhysRevB.106.245145}{Phys. Rev. B \textbf{106}, 245145 (2022)}.

\bibitem{CeSbTe_Lv} B. Lv, J. Chen, L. Qiao, J. Ma, X. Yang, M. Li, M. Wang, Q. Tao, and Z.-A. Xu, Magnetic and transport properties of low-carrier-density Kondo semimetal CeSbTe, \href{https://dx.doi.org/10.1088/1361-648X/ab2498}{J. Phys. Condens. Matter \textbf{31}, 355601 (2019)}.

\bibitem{CeSbTe_peng} P. Li, B. Lv, Y. Fang, W. Guo, Z. Wu, Y. Wu, D. Shen, Y. Nie, L. Petaccia, C. Cao, Z.-A. Xu, and Y. Liu, Charge density wave and weak Kondo effect in a Dirac semimetal $\mathrm{CeSbTe}$, \href{https://doi.org/10.1007/s11433-020-1642-2}{Sci. China: Phys. Mech. Astron. \textbf{64}, 237412 (2021)}.

\bibitem{CeSbSe_Chen} K.-W. Chen, Y. Lai, Y.-C. Chiu, S. Steven, T. Besara, D. Graf, T. Siegrist, T. E. Albrecht-Schmitt, L. Balicas, R. E. Baumbach, Possible devil's staircase in the Kondo lattice CeSbSe, \href{https://link.aps.org/doi/10.1103/PhysRevB.96.014421}{Phys. Rev. B \textbf{96}, 014421 (2017)}.

\bibitem{SmSbTe_sabin} S. Regmi, G. Dhakal, F. C. Kabeer, N. Harrison, F. Kabir, A. P. Sakhya, K. Gofryk, D. Kaczorowski, P. M. Oppeneer, M. Neupane, Observation of multiple nodal lines in SmSbTe, \href{https://link.aps.org/doi/10.1103/PhysRevMaterials.6.L031201}{Phys. Rev. Mater. \textbf{6}, L031201 (2022)}.

\bibitem{PrSbTe} S. Regmi, I. B. Elius, A. P. Sakhya, M. Sprague, M. I. Mondal, N. Valadez, V. Buturlim, K. Booth, T. Romanova, K. Gofryk, A. Ptok, D. Kaczorowski, M. Neupane, Electronic structure in a rare-earth based nodal-line semimetal candidate PrSbTe, \href{https://link.aps.org/doi/10.1103/PhysRevMaterials.8.L041201}{Phys. Rev. Mater. \textbf{8}, L041201 (2024)}.

\bibitem{NdSbTeEandMProp_Pandey} K. Pandey, R. Basnet, A. Wegner, G. Acharya, M. R. U. Nabi, J. Liu, J. Wang, Y. K. Takahashi, B. Da, J. Hu, Electronic and magnetic properties of the topological semimetal candidate NdSbTe, \href{https://link.aps.org/doi/10.1103/PhysRevB.101.235161}{Phys. Rev. B \textbf{101}, 235161 (2020)}.

\bibitem{Nd111} S. Regmi, R. Smith, A. P. Sakhya, M. Sprague, M. I. Mondal, I. B. Elius, N. Valadez, A. Ptok, D. Kaczorowski, M. Neupane, Observation of gapless nodal-line states in NdSbTe, \href{https://link.aps.org/doi/10.1103/PhysRevMaterials.7.044202}{Phys. Rev. Mater. \textbf{7}, 044202 (2023)}.

\bibitem{GdSbTe_hosen} M. M. Hosen, G. Dhakal, K. Dimitri, P. Maldonado, A. Aperis, F. Kabir, C. Sims, P. Riseborough, P. M. Oppeneer, D. Kaczorowski, T. Durakiewicz, M. Neupane, Discovery of topological nodal-line fermionic phase in a magnetic material GdSbTe, \href{https://doi.org/10.1038/s41598-018-31296-7}{Sci. Rep. \textbf{8}, 13283 (2018)}.

\bibitem{HoSbTe_arpes} S. Yue, Y. Qian, M. Yang, D. Geng, C. Yi, S. Kumar, K. Shimada, P. Cheng, L. Chen, Z. Wang, H. Weng, Y. Shi, K. Wu, B. Feng, Topological electronic structure in the antiferromagnet HoSbTe, \href{https://link.aps.org/doi/10.1103/PhysRevB.102.155109}{Phys. Rev. B \textbf{102}, 155109 (2020)}.

\bibitem{HoSbTe_shumiya} N. Shumiya, J.-X. Yin, G. Chang, M. Yang, S. Mardanya, T.-R. Chang, H. Lin, M. S. Hossain, Y.-X. Jiang, T. A. Cochran, Q. Zhang, X. P. Yang, Y. Shi, M. Z. Hasan, Evidence for electronic signature of a magnetic transition in the topological magnet HoSbTe, \href{https://link.aps.org/doi/10.1103/PhysRevB.106.035151}{Phys. Rev. B \textbf{106}, 035151 (2022)}.

\bibitem{HoSBTe_yang} M. Yang, Y. Qian, D. Yan, Y. Li, Y. Song, Z. Wang, C. Yi, H. L. Feng, H. Weng, Y. Shi, Magnetic and electronic properties of a topological nodal line semimetal candidate: HoSbTe, \href{https://link.aps.org/doi/10.1103/PhysRevMaterials.4.094203}{Phys. Rev. Mater. \textbf{4}, 094203 (2020)}.

\bibitem{LaSbTe_transport} K. Pandey, L. Sayler, R. Basnet, J. Sakon, F. Wang, J. Hu, Crystal Growth and Electronic Properties of LaSbSe, \href{https://www.mdpi.com/2073-4352/12/11/1663}{Crystals \textbf{12}, 1663 (2022)}.

\bibitem{Schoop} L. M. Schoop, A. Topp, J. Lippmann, F. Orlandi, L. Müchler, M. G. Vergniory, Y. Sun, A. W. Rost, V. Duppel, M. Krivenkov, S. Sheoran, P. Manuel, A. Varykhalov, B. Yan, R. K. Kremer, C. R. Ast, B. V. Lotsch, Tunable Weyl and Dirac states in the nonsymmorphic compound CeSbTe, \href{http://doi.org/10.1126/sciadv.aar2317}{Sci. Adv. \textbf{4}, eaar2317 (2018)}.

\bibitem{Gao_DySbTe} F. Gao, J. Huang, W. Ren, M. Li, H. Wang, T. Yang, B. Li, Z. Zhang, Magnetic and transport properties of the topological compound DySbTe, \href{https://link.aps.org/doi/10.1103/PhysRevB.105.214434}{Phys. Rev. B \textbf{105}, 214434 (2022)}.

\bibitem{plokhikh_Ln} I. Plokhikh, V. Pomjakushin, D. J. Gawryluk, O. Zaharko, E. Pomjakushina, On the magnetic structures of 1:1:1 stoichiometric topological phases $Ln$SbTe ($Ln$= Pr, Nd, Dy, and Er), \href{https://doi.org/10.1016/j.jmmm.2023.171009}{J. Magn. Magn. Mater. \textbf{583}, 171009 (2023)}.

\bibitem{SupplementalMat} See Supplemental Material at (Link) for experimental and computational details, photon-dependent measurements, and additional details on STM.

\bibitem{kresse.hafner.94} G. Kresse, J. Hafner, Ab initio molecular-dynamics simulation of the liquid-metal--amorphous-semiconductor transition in germanium, \href{http://doi.org/10.1103/PhysRevB.49.14251}{Phys. Rev. B \textbf{49}, 14251 (1994)}.

\bibitem{kresse.furthmuller.96} G. Kresse, J. Furthmüller, Efficient iterative schemes for ab initio total-energy calculations using a plane-wave basis set, \href{http://doi.org/10.1103/PhysRevB.54.11169}{Phys. Rev. B \textbf{54}, 11169 (1996)}.

\bibitem{kresse.joubert.99} G. Kresse, D. Joubert, From ultrasoft pseudopotentials to the projector augmented-wave method, \href{http://doi.org/10.1103/PhysRevB.59.1758}{Phys. Rev. B \textbf{59}, 1758 (1999)}.

\bibitem{blochl.94} P. E. Blöchl, Projector augmented-wave method, \href{http://doi.org/10.1103/PhysRevB.50.17953}{Phys. Rev. B \textbf{50}, 17953 (1994)}.

\bibitem{Apiezon} W. Schnelle, J. Engelhardt, E. Gmelin, \href{https://doi.org/10.1016/S0011-2275(99)00035-1}{Cryogenics \textbf{39}, 271 (1999)}.

\bibitem{nonstoichiometry} I. Plokhikh, O. Fabelo, L. Prodan, M. Wörle, E. Pomjakushina, A. Cervellino, V. Tsurkan, I. Kézsmárki, O. Zaharko, \href{https://doi.org/10.1016/j.jallcom.2022.168348}{J. Alloys Compd. \textbf{936}, 168348 (2023)}.

\bibitem{bian_visualizing_2022} Q. Bian, S. Li, A. Luo, Z. Zhang, J. Hu, Y. Zhu, Z. Shao, H. Sun, Z. Cheng, Z. Mao, G. Xu, M. Pan, Visualizing discrete Fermi surfaces and possible nodal-line to Weyl state evolution in ZrSiTe, \href{https://www.nature.com/articles/s41535-022-00463-5}{npj Quantum Mater. \textbf{7}, 1–8 (2022)}.
\end{thebibliography}
\end{document}